\documentclass[twocolumn,showpacs,preprintnumbers,amsmath,amssymb]{revtex4}
\usepackage{graphicx}

\begin{document}

\title{Direct bandgap optical transitions in Si nanocrystals}
\author{A.~A.~Prokofiev$^1$,
A.~S.~Moskalenko$^{2,1}$,
I.~N.~Yassievich$^1$}
\affiliation{$^1$Ioffe Physico-Technical Institute,
Polytechnicheskaya 26, 194021 Saint-Petersburg, Russia
\\
$^2$Institut f\"ur Physik, Martin-Luther-Universit\"at
Halle-Wittenberg, Germany}
\author{W.~D.~A.~M.~de Boer$^3$,
D.~Timmerman$^3$,
H.~Zhang$^4$,
W.~J.~Buma$^4$ and
T.~Gregorkiewicz$^3$} \affiliation{$^3$Van der
Waals-Zeeman Institute, University of Amsterdam,\\
Valckenierstraat 65, NL-1018 XE Amsterdam, The Netherlands\\
$^4$Van~'t Hoff Institute for Molecular Sciences, University of Amsterdam\\
Nieuwe Achtergracht 129, NL-1018 WS Amsterdam, The Netherlands}
\date{\today}

\begin{abstract}
The effect of quantum confinement on the direct bandgap of
spherical Si nanocrystals has been modelled theoretically.  We
conclude that the energy of the direct bandgap at the
$\Gamma$-point decreases with size reduction: quantum confinement
enhances radiative recombination across the direct bandgap and
introduces its ``red" shift for smaller grains. We postulate to
identify the frequently reported efficient blue emission
(\textit{F}-band) from Si nanocrystals with this zero-phonon
recombination. In a dedicated experiment, we confirm the ``red"
shift of the \textit{F}-band, supporting the proposed
identification.
\end{abstract}

\maketitle
Crystalline silicon, which dominates electronic and photovoltaic
industry, has notoriously inferior optical properties for light
emission. This follows from its energy structure which features an
indirect bandgap, with the absolute minimum of the conduction band
being very much displaced --- $\Delta k = 0.85 \times (2 \pi/a)$
along the [100]-direction --- from the $\Gamma$-point, where the
maximum of the valence band is positioned. The direct bandgap
energy at the $\Gamma$-point amounts to
$E_{\mathrm{g}}^{\mathrm{dir}}$~=~3.32 eV in bulk Si. On the way
towards photonic and opto-electronic applications of silicon,
possibly the most promising route is offered by
space-confinement-induced changes of energy structure. Following
the pioneering paper of Canham on porous Si \cite{Po-Si},
investigations of various forms of nanostructured Si have been
undertaken. In particular, very interesting results were obtained
for Si nanocrystals (NCs) where strong effects of quantum
confinement have been observed for grain sizes comparable to or
smaller than the effective Bohr radius of 4.3 nm --- e.g. see
Ref.~\cite{Kovalev-rev} for an extensive review. It has been shown
that while the energy structure in Si NCs retains its indirect
character, quantum confinement leads to relaxation of the momentum
conservation requirement, thus making zero-phonon optical
interband transitions possible. Nevertheless, phonon-assisted
radiative transitions dominate for nanocrystal sizes down to
2~nm~\cite{Kovalev}. Therefore, the emission from Si-NCs has two
characteristic features: for smaller grains, the spectrum shifts
to the blue and its intensity increases. In the literature, this
widely tuneable and rather efficient emission due to ground state
electron-hole recombination is usually referred to as the ``slow"
band (\textit{S}-band). It has attracted a lot of attention and
prospects of Si-NCs-based laser have been discussed~\cite{Pavesi}.

In addition to the $S$-band whose origin is well established,
optical investigations of Si-NCs revealed also another emission
band at higher energies. This ``blue" band is characterized by
much faster decay dynamics (hence the name \textit{F}-band),
ranging from pico- to nanoseconds. A similar emission band has
also been reported for porous Si~\cite{Tsy-RC}. Its origin is
currently debated both experimentally and theoretically
\cite{Trojanek, Kuntermann, 3}. Usually, it is postulated to arise
due to radiative recombination of electrons localized at
oxygen-related states at Si/SiO$_2$ interface ~\cite{Wolkin,
Delerue}. This interpretation has been recently challenged by
Valenta~\cite{JV}, who presented evidence that the \textit{F}-band
should rather be identified with transitions related to the NCs
themselves. Emission bands in the visible have also been reported
in colloidal Si-NCs \cite{Wilcoxon, English} and postulated to
arise due to direct transitions in view of their short lifetime.

In this paper, we consider effect of quantum confinement on the
conduction band at the $\Gamma$-point in spherically symmetric Si-NCs.
According to the presented theoretical model, quantum confinement
applied to the conduction band states in the vicinity of
the $\Gamma$-point, creates two series of states with energies higher
and lower than the energy of the conduction band at the $\Gamma$-point
in bulk Si (3.32~eV), forming a gap between them. We attribute the
lower series of states as the origin of direct optical transitions
responsible for the \textit{F}-band. In that interpretation, the
\textit{F}-band is microscopically identified as arising due to
zero-phonon recombination of ``hot'' electrons from the $\Gamma$-point of the conduction band. We discuss that
as the confinement gets stronger with decreasing the NC size and
the direct band gap shrinks, the ``red'' shift of the
\textit{F}-band appears. We confirm this theoretical conclusion by
dedicated experimental evidence concerning spectral dependence and
dynamics of the \textit{F}-band in a dense solid state dispersion
of Si-NCs in a SiO$_2$ matrix. We demonstrate that the proposed
microscopic identification of the \textit{F}-band is consistent
with this experimental data as well as with the available
literature.

\begin{figure}
\includegraphics[width=8cm]{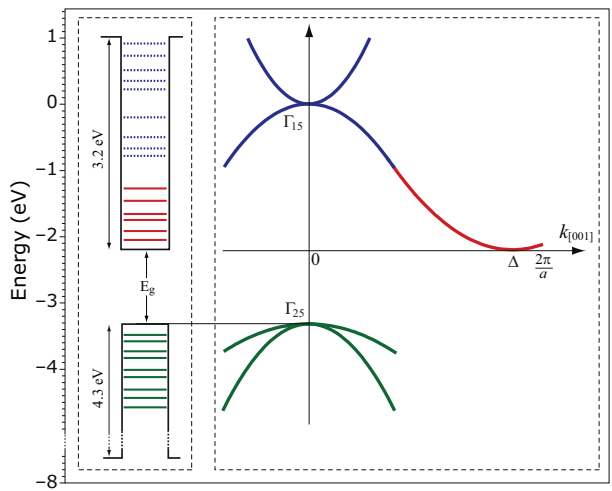}
\caption{Fig.~1. Right panel: schematical representation of the
dispersion relations for the conduction and the valence bands of
bulk silicon along the [001] direction~\cite{Cardona_PR_1966}.
Left panel: quantum potentials in real space and confinement
energy levels for electrons and holes. Our model attributes the
lower quantum levels (solid lines) of electrons to the states originating from
$\Delta$-valley of the conduction band, and the higher excited
ones (dotted lines) to the states of the $\Gamma$-point.}
\end{figure}

The scheme of the lowest conduction band in bulk silicon (based on
detailed calculations from Ref.~\cite{Cardona_PR_1966}) is shown
in the right panel of Fig.~1. In the center of Brillouin zone its
states are of $\Gamma_{15}$ symmetry, thus being 3-fold degenerate
(without spin taken into account), and they are related to two
subbands, one of which (the ``heavy'' one) has negative effective
mass. It is the negative effective mass that leads
to emerging
of space-confinement levels below the band edge and therefore to
decrease of the threshold of direct optical transitions.

Counting energy of electrons from the $\Gamma_{15}$-point, one can
write a generalization of the Luttinger Hamiltonian in spherical
approximation:
\begin{equation}
    \hat{H}
      = (A+2B) \frac{\hat{\mathbf{p}}^2}{m_0}
      - 3 B
        \frac{(\hat{\mathbf{p}}\cdot\hat{\mathbf{J}})^2}{m_0},
\label{ham}
\end{equation}
where $m_0$ is the free electron mass, $\hat{\mathbf{p}}$ is the
momentum operator, and $\hat{\mathbf{J}}$ is the unitary angular
momentum operator acting in the space of Bloch amplitudes. The
coefficients $A$ and $B$, related to the effective mass values,
are extracted from the results of Ref.~\cite{Cardona_PR_1966} by
weightening the band parameters obtained for different directions:
$A = 0.4 A_{001} + 0.6 A_{111} =  0.58$, $B = 0.4 B_{001} + 0.6
B_{111} = -0.85$. Such values correspond to positive energy and
``light'' effective mass of $m_{el}=0.35 \, m_0$ for doubly
degenerate subband and negative energy and ``heavy'' effective
mass of $m_{eh}=-0.45 \, m_0$ for the other nondegenerate subband.
In a spherical quantum dot Hamiltonian~(\ref{ham}) results in
three types of states~\cite{Moskalenko_PRB_2007}, which are the
eigenfunctions of the square $\hat{F^{2}}$ of the full angular
momentum operator $\hat{\bf F}=\hat{\bf J}+\hat{{\bf L}}$
(operator $\hat{{\bf L}}=-i{\bf r}\times {\nabla}$ acts on the
envelope parts of the wave functions only) and its projection
$\hat{F_{z}}$. They are

i) ``heavy'' states $Neh_0$ with $F=0$,

ii) ``light'' states $Nel_F$ ($F\geq 1$),

iii) ``mixed'' states  $\pm Nem_F$ ($F\geq 1$),

\noindent where $N$ is the main quantum number
and the subscript shows the value of $F$.

\begin{figure}
\includegraphics[width=8cm]{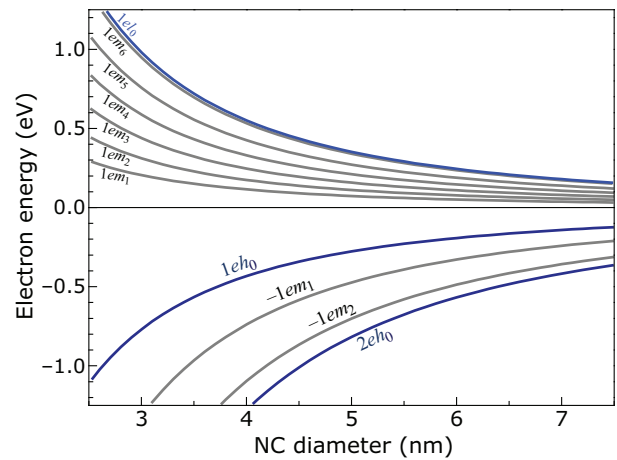}
\caption{Fig.~2. Energy levels of electrons at the $\Gamma$-point in
spherical Si-NC as a function of its diameter. According to
Fig.~1, zero energy corresponds to the energy of bulk Si
conduction band at the $\Gamma$-point.}
\end{figure}

To investigate qualitatively the effect of confinement on the
energy spectrum, we calculated the space-quantization levels of
electrons in the vicinity of the $\Gamma_{15}$-point assuming for
them infinitely high energy barriers at the NC boundary. The
resulting energy spectrum as a function of NC size is shown in
Fig.~2 . One can see that the spectrum is split: ``light'' states
have positive energy, ``heavy'' --- negative, and ``mixed'' ones
have both positive and negative energy levels. The most important
feature of the resulting spectrum is that the energy gap is formed
and  the absolute values of negative energy levels are rising with
decreasing diameter of NCs. We should remark that the appearance
of such a gap was predicted in Ref.~\cite{Rama_Krishna} with help
of rudimentary pseudopotential calculations.  The quantization
energies of holes are rising as well~\cite{Moskalenko_PRB_2007},
but it does not change the trend, and the transition energy
does decrease for given
transition with decreasing the size.

The rate of direct spontaneous optical recombination is given by~\cite{Delerue_book_2004}:
\begin{equation}
    \tau_{\mathrm{rad}}^{-1}
    = \frac{4}{3}
      \alpha^3
      \omega
      n_{\mathrm{out}}
      |M|^2
      \mathcal{F}^2,
\label{tau-1}
\end{equation}
where  $\alpha=\hbar/(m_0 a_0 c) \approx 1/137$ is the fine
structure constant, $a_0$ is Bohr radius, $c$ is the speed of
light, $\hbar\omega$ is the transition energy, $\mathcal{F}= 3
n^2_{\mathrm{out}}/ (n^2_{\mathrm{in}} + 2 n^2_{\mathrm{out}})$ is
the local field factor~\cite{Delerue_book_2004}, with
$n_{\mathrm{in}}$ and $n_{\mathrm{out}}$ being refraction indexes
of media inside and outside the NC respectively (for the NCs under
consideration $\mathcal{F}\approx 0.42$). The dimensionless matrix
element $M$ in~(\ref{tau-1}) is defined by:
\begin{multline}
|M|^2 =\frac{a_{0}^2}{\hbar^2}
    \frac{1}{N_{\rm e}}
    \frac{1}{N_{\rm h}}
    \sum_{M_{\rm e},M_{\rm h}}
\\
    \sum_{\alpha=x,y,z}
    \left|
        \sum_{v,c}
            \int \left(\psi_{v}^{M_{\rm h}}\right)^* \psi_{c}^{M_{\rm e}} d^3r
            \langle
                u_{v} | \hat{p_\alpha} | u_{c}
            \rangle
    \right|^2,
\label{M2}
\end{multline}
where
$N_{\rm e}=2F_{\rm e}+1$ and $N_{\rm h}=2F_{\rm h}+1$ are the degeneracy factors of the electron and hole states respectively,
and $\psi_{c}^{M_{\rm e}}$ ($\psi_{v}^{M_{\rm h}}$) are the envelope parts of the electron (hole) wave function
$\Psi_{F_{\rm e}M_{\rm e}}=\sum_c \psi_{c}^{M_{\rm e}} u_c$ ($\Psi_{F_{\rm h}M_{\rm h}}=\sum_v \psi_{v}^{M_{\rm h}} u_v$).
The Bloch amplitudes basis $u_{c}$ is chosen in the
form of spherical components $u_0=Z_c$,
$u_\pm=\mp\sqrt{1/2}(X_c\pm iY_c)$, where $X_c=x$, $Y_c=y$, and
$Z_c=z$ are the functions of the $\Gamma_{15}$ representation.
It should be noted that the consideration of hole states
in~\cite{Moskalenko_PRB_2007} has also been produced with the
Bloch amplitude basis $u_v$ in the form of spherical components but
using functions $X_v = yz$, $Y_v = xz$, and $Z_v = xy$ of the
$\Gamma_{25'}$ representation.
The difference in the Bloch
amplitude bases makes the direct optical transitions between
electron and hole states permitted.

The value of the matrix element of momentum between the Bloch
amplitudes of the valence and the conduction bands $p_{cv}$
allowed by the selection rules can be written in the form:
$p^2_{cv} = \left|
                \langle
                    X_c | \hat{p}_y | Z_v
                \rangle
            \right|^2
            =
            \left|
                \langle
                    Y_c | \hat{p}_z | X_v
                \rangle
            \right|^2
            =
            \left|
                \langle
                    Z_c | \hat{p}_x | Y_v
                \rangle
            \right|^2
            =
            Q^2 {\hbar^2}/{a_0^2}$,
where $Q=1.07$~\cite{Cardona_PR_1966}.
And the total matrix element~(\ref{M2}) is $|M|^2
= Q^2 I_R^2$, where $I_R$ is a combination of overlap integrals
of the envelope functions. There are also selection rules for overlap integrals of envelope
functions in Eq.~(\ref{M2}) arising due to the spherical symmetry
of both electron and hole states. E.g., in our consideration,
radiative transitions from ``heavy'' electron states $Neh_0$ can
only go into the valence band states of ``mixed'' type,
characterized by $F=2$: $Nhm_2$. These rules should not be strict
for real NCs, which are not perfectly spherical, and transitions
disabled by our model are still possible, but we expect them to be
much weaker.

\begin{table}
\caption{
    \label{tab:I_R}
    Factors $I_R^2$ for several interband transitions, as well as their energies for two NC sizes: 3 and 7~nm.}
        \begin{tabular}{c|cc|cc}
             & \multicolumn{2}{c|}{$d=3$~nm} & \multicolumn{2}{c}{$d=7$~nm}
            \\
            Transition & $I_R^2$ & $\hbar\omega$ & $I_R^2$ & $\hbar\omega$
            \\
            \hline
            $\phantom{-}1eh_0 \rightarrow 1hm_2$ & 1.185 & 2.95 eV & 1.277 & 3.28 eV
            \\
            $-1em_1 \rightarrow 1hh_2$           & 0.099 & 2.53 eV & 0.130 & 3.23 eV
            \\
            $-1em_1 \rightarrow 1hm_1$           & 0.016 & 2.17 eV & 0.002 & 3.13 eV
            \\
            $\phantom{-}2eh_0 \rightarrow 1hm_2$ & 0.077 & 1.45 eV & 0.047 & 3.01 eV
        \end{tabular}
\end{table}

Finally, for the estimation of radiative recombination rate one
can use:
\begin{equation}
    \tau^{-1}_{\rm rad} = 1.0 \times 10^{8} \left(\frac{\hbar\omega}{3\,\mathrm{eV}}\right) I_R^2,\,\mathrm{s}^{-1}.
\label{tau_number}
\end{equation}
The values of $I_R^2$ for the radiative transitions with the
smallest energies are listed in
Table~\ref{tab:I_R}.
For the most effective
transition $Neh_0 \rightarrow Nhm_2$ the radiative life time is of
the order of 10~ns. We propose that the emissions observed as the
\textit{F}-band arise due to transitions in Table~\ref{tab:I_R}.
Table~\ref{tab:I_R} also illustrates that the energy range of
possible transitions increases with decrease of NC diameter.

The radiative emission under consideration can be produced only by
``hot'' confined electrons. Thus the rate of nonradiative
relaxation of these electrons is a key issue for the observation
of this emission. We have studied transitions between levels in
the vicinity of the $\Gamma_{15}$-point taking place due to
multiple emission of optical phonons and  emission/absorption of
one promoting acoustic longitudinal phonon. The probability of
such a transition is given by $W=\tau_{\rm ac}^{-1} J_{p}(T,S)$, where $\tau_{\rm ac}$ is the time determined by
the interaction with an acoustic phonon, $T$ is temperature, $S$
is the Huang-Rhys factor, and $J_{p}(T,S)$ is the factor specific
for multiphonon transitions which for $S \ll 1$ and $kT$ less than
the optical phonon energy $\hbar \omega\approx 60$~meV can be well
approximated by $S^{p}/p!\:$, where $p$ is the number of emitted
optical phonons \cite{Goupalov}. We have calculated the Huang-Rhys
factor following Ref.~\cite{Goupalov} and using Bir-Pikus
Hamiltonian $    H_{\rm e-ph}^{\rm opt}({\bf u}^{\rm opt})
    = \frac{2}{\sqrt{3}} \frac{d_0}{a}
    \left(
        u_x^{\rm opt} \, \{J_y J_z\} +u_y^{\rm opt} \{J_z J_x\} +u_z^{\rm opt} \{J_x J_y\}
    \right)$ for the interaction with optical phonons,
where $2\{J_\alpha J_\beta \}\equiv J_\alpha J_\beta + J_\beta
J_\alpha $, $a$ is the lattice constant, $d_0=-16.9$~eV
\cite{Blacha} is the interaction constant, and ${\bf u}^{\rm opt}$
is the relative atomic displacement induced by the optical phonon
mode. In result we have $S=0.13/d^3$ for $1eh_0\!\rightarrow\!
-1em_1$ transition, $S=0.04/d^3$ for $-1em_1\!\rightarrow\!
-1em_2$, and $S=0.22/d^3$ for $-1em_2\!\rightarrow\! 2eh_0$, where
the NC diameter $d$ should be taken in nm. Further we have
calculated $\tau_{\rm ac}^{-1}$ and then $W$ taking into account
the energy difference which is not compensated by the optical
phonons and using the interaction Hamiltonian $H_{\rm e-ph}^{\rm
ac}=a_c \nabla {\bf u}^{\rm ac}$, where $a_c=-10$~eV \cite{Blacha}
and ${\bf u}^{\rm ac}$ is the atomic displacement induced by the
acoustic phonon mode. The highest transition rates are of the
order of $10^9$~s$^{-1}$ for the largest  NCs considered here.
They decrease rapidly with the NC size. One can notice that for
transitions $1eh_0\!\rightarrow\! -1em_1$ and
$-1em_1\!\rightarrow\! -1em_2$ several optical phonons are needed
even for large NCs with extremely small Huang-Rhys factors (see
Fig.~2). In case of small NCs more phonons are required and
Huang-Rhys factors are still small so that the phonon-induced
relaxation is suppressed. The state $-1em_2$ can decay quicker
than the other states due to smaller energy spacing to the lower
neighboring state but it anyway does not contribute to the
considered radiative transitions (see Table I).

\begin{figure}
\includegraphics[width=8cm]{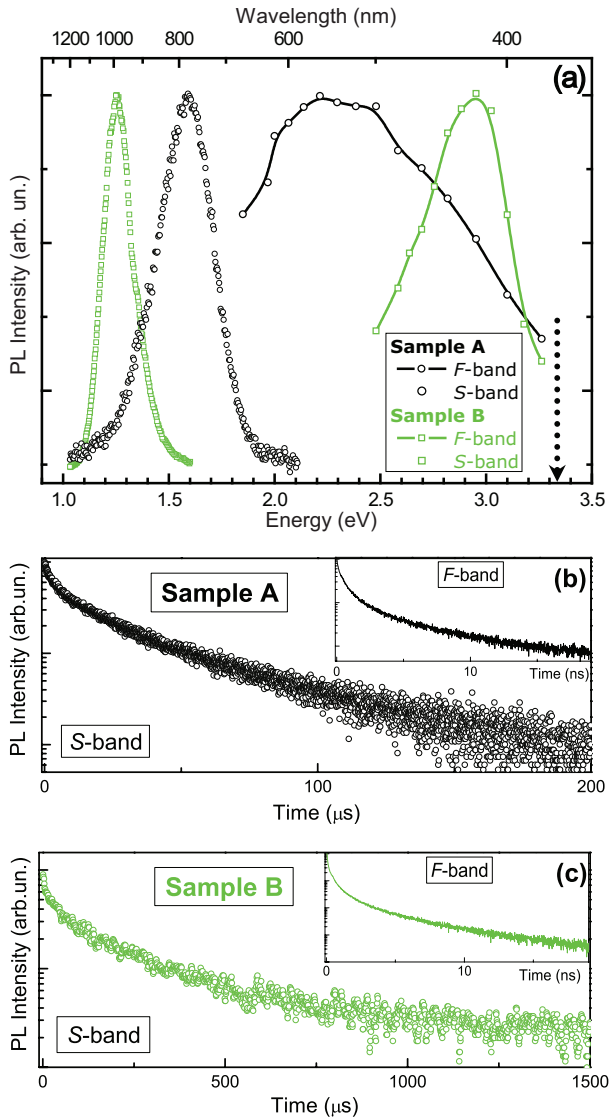}
\caption{Fig.~3. (a) \textit{S} (scattered symbols) and
\textit{F}(solid line + symbols) PL bands for sample A (circles) and sample B
(squares). Dashed arrow indicates the
position of direct bandgap recombination as experimentally
measured for bulk Si \cite{STM}. (b,c) PL dynamics for sample A
(figure b) and sample B (figure c): \textit{S}-band decay is
displayed in main panels. The kinetics were fitted with a
stretched exponential:
I$_{\mathrm{PL}}$=I$_{0}$e$^{-(t/\tau)^{\beta}}$, with $\beta$ =
0.8 \cite{Linross}. The relevant time constants are
$\tau_{\mathrm{A}}^{\mathrm{S}} \approx 20$~$\mu$s for sample A
and $\tau_{\mathrm{B}}^{\mathrm{S}} \approx 510$~$\mu$s for sample
B. The inset shows dynamics of the \textit{F}-band in a time
window of 50~ns taken at the maximum intensity values of 2.5~eV
and 3~eV for samples A and B, respectively.}
\end{figure}

The proposed theoretical identification of the \textit{F}-band is
directly supported by photoluminescence (PL) experiments performed
on Si-NCs embedded in a SiO$_2$ matrix. Two different samples were
used, both prepared by radio frequency co-sputtering and
subsequent annealing, resulting in an average NC diameter of 3 nm
(sample A) and 5.5 nm (sample B), with a small size distribution
--- see, e.g. Refs.~\cite{Fujii, NP} for more details on sample
preparation and measuring techniques. Figure 3 shows the results
obtained for both samples in two differently conducted PL
experiments. With the steady-state technique, the slow
(microsecond range) time-integrated PL signal was monitored. This
is dominated by the low-energy \textit{S}-band (indicated in Fig.
3(a) with ~$\put (0,3) {\circle{6}}$~~) and, as a result of
quantum confinement, shows the well-known shift towards higher
photon energies for smaller NCs. The spectral profile of the
\textit{F}-band has been obtained by taking the initial amplitude
of the ps-resolved dynamics and is illustrated in Fig. 3(a)(---).
As can be seen, the \textit{F}-band shows a red shift upon size
decrease --- an effect opposite to the blue shift characteristic
for the \textit{S}-band. This behavior is in agreement with the
presently calculated shift towards lower energies of the electron energy
levels at the $\Gamma$-point
--- see Fig.~2 --- and the consequent reduction of the direct bandgap
value. We note that the linewidth of the \textit{F}-band clearly
increases for smaller NCs, as indeed expected due to quantum
confinement. The time-resolved PL measurements have been obtained
in a time-correlated single photon counting experiment. Figs 3(b)
and 3(c) illustrate dynamics of \textit{S} and \textit{F} PL bands
for both samples. The decay time of the \textit{S}-band (main
panels, nanosecond resolution) is of an order of $\sim10^2~\mu$s,
i.e. similar to the estimated radiative recombination time of
excitons in Si-NCs \cite{Moskalenko_PRB_2007,Watanabe_JAP_2001}.
It is somewhat faster for smaller NCs (sample A in respect to B),
reflecting the increased radiative recombination rate~\cite{Vic}.
The decay dynamics of the \textit{F}-band (insets, picosecond
resolution) is characterized by a time constant of
$\tau_\mathrm{F} \approx 500$ ps, similar for both samples.
Finally, we point out that the fact that the \textit{F}-band can
be monitored using (ps-resolution) time-resolved PL experiments,
implies that an upper limit can be set to the ratio of the
non-radiative transition rate to the radiative recombination rate
of
$(\tau_{\mathrm{nonrad}}^{\mathrm{F}})^{-1}\div(\tau_{\mathrm{rad}}^{\mathrm{F}})^{-1}\leq10^2$.
On that basis, the radiative recombination time for the
\textit{F}-band can be estimated as
$\tau^{\mathrm{F}}_{\mathrm{rad}}\leq50$~ns.

We conclude that the characteristic features following from the
proposed theoretical model for the direct $\Gamma$-point
phonon-less recombinations
--- (i) decrease of the energy and (ii) linewidth broadening for
smaller NC sizes are experimentally confirmed for the
\textit{F}-band emission from Si-NCs. Also the radiative lifetime
of the \textit{F}-band estimated from the measured PL dynamics
agrees with the calculated values. These additional new
experimental evidence taken together with the information on
\textit{F}-band available in the literature, agree very well with
the proposed identification of this fast emission with radiative
recombination across the direct bandgap of Si, whose energy has
been lowered in Si-NCs due to quantum confinement.

Further research, currently on the way, will elucidate details of
the direct bandgap tuning in Si-NCs.

This work has been supported by Stichting voor de Technologische
Wetenschappen (STW), Nederlandse Organisatie voor Wetenschappelijk
Onderzoek (NWO), and Russian Foundation for Basic Research (RFBR)
as well as grants of the Russian President. The samples on which
the experimental results were conducted, have been developed in
cooperation with Dr. M. Fujii from Kobe University.

\end{document}